\documentstyle{article}
\input math_macros
\font\tenbf=cmbx10

\font\tenrm=cmr10
\font\tenit=cmti10
\font\elevenbf=cmbx10 scaled\magstep 1
\font\elevenrm=cmr10 scaled\magstep 1
\font\elevenit=cmti10 scaled\magstep 1

\font\ninerm=cmr9

\def \cp89{{\it CP Violation,} edited by C. Jarlskog (World Scientific,
Singapore, 1989)}

\def \hb87{{\it Proceeding of the 1987 International Symposium on Lepton and
Photon Interactions at High Energies,} Hamburg, 1987, ed. by W. Bartel
and R. R\"uckl (Nucl. Phys. B, Proc. Suppl., vol. 3) (North-Holland,
Amsterdam, 1988)}

\def \ichep72{{\it Proceedings of the XVI International Conference on High
Energy Physics}, Chicago and Batavia, Illinois, Sept. 6 -- 13, 1972,
edited by J. D. Jackson, A. Roberts, and R. Donaldson (Fermilab, Batavia,
IL, 1972)}

\def \ite{{\it et al.}}
\def \lkl87{{\it Selected Topics in Electroweak Interactions} (Proceedings of
the Second Lake Louise Institute on New Frontiers in Particle Physics, 15 --
21 February, 1987), edited by J. M. Cameron \ite~(World Scientific, Singapore,
1987)}
\def \ky85{{\it Proceedings of the International Symposium on Lepton and
Photon Interactions at High Energy,} Kyoto, Aug.~19-24, 1985, edited by M.
Konuma and K. Takahashi (Kyoto Univ., Kyoto, 1985)}

\def \si90{25th International Conference on High Energy Physics, Singapore,
Aug. 2-8, 1990}
\def \slc87{{\it Proceedings of the Salt Lake City Meeting} (Division of
Particles and Fields, American Physical Society, Salt Lake City, Utah, 1987),
ed. by C. DeTar and J. S. Ball (World Scientific, Singapore, 1987)}
\def \slac89{{\it Proceedings of the XIVth International Symposium on
Lepton and Photon Interactions,} Stanford, California, 1989, edited by M.
Riordan (World Scientific, Singapore, 1990)}
\def \smass82{{\it Proceedings of the 1982 DPF Summer Study on Elementary
Particle Physics and Future Facilities}, Snowmass, Colorado, edited by R.
Donaldson, R. Gustafson, and F. Paige (World Scientific, Singapore, 1982)}
\def \smass90{{\it Research Directions for the Decade} (Proceedings of the
1990 Summer Study on High Energy Physics, June 25--July 13, Snowmass,
Colorado),
edited by E. L. Berger (World Scientific, Singapore, 1992)}
\def \tasi90{{\it Testing the Standard Model} (Proceedings of the 1990
Theoretical Advanced Study Institute in Elementary Particle Physics, Boulder,
Colorado, 3--27 June, 1990), edited by M. Cveti\v{c} and P. Langacker
(World Scientific, Singapore, 1991)}

\renewenvironment{thebibliography}[1]
 { \elevenrm
   \begin{list}{\arabic{enumi}.}
    {\usecounter{enumi} \setlength{\parsep}{0pt}
     \setlength{\itemsep}{3pt} \settowidth{\labelwidth}{#1.}
     \sloppy
    }}{\end{list}}

\begin{document}
\begin{center}{{\tenbf TEST FOR RIGHT-HANDED $b$ QUARK DECAYS\footnote{
\ninerm Presented by Jonathan L. Rosner at DPF 92 Meeting, Fermilab,
November, 1992.}\\}
\vspace{-1in}
\rightline{EFI 92-63}
\rightline{November 1992}
\bigskip
\vglue 2.0cm
{\tenrm JAMES F. AMUNDSON, JONATHAN L. ROSNER, AND MIHIR P. WORAH\\}
\baselineskip=13pt
{\tenit Enrico Fermi Institute and Department of Physics,
University of Chicago\\}
\baselineskip=12pt
{\tenit 5640 S. Ellis Ave., Chicago, IL 60637, USA\\}
 \vglue 0.3cm
 {\tenrm and\\}
 \vglue 0.3cm
 {\tenrm MARK B. WISE\\}
 {\tenit Lauritsen Laboratory of Physics, California Institute of
Technology\\}
 {\tenit Pasadena, CA 91125 \\}
\vglue 0.8cm
{\tenrm ABSTRACT}}
\end{center}
\vglue 0.3cm
{\rightskip=3pc
 \leftskip=3pc
 \tenrm\baselineskip=12pt
 \noindent
Gronau and Wakaizumi have proposed a model in which the dominant $b$ decays are
due to exchange of a new right-handed gauge boson.  A test of this model via
the study of polarized $\Lambda_b$ baryons produced in $e^+ e^- \to Z \to
\Lambda_b + X$ is suggested.
\vglue 0.6cm}
\vglue 0.4cm
\baselineskip=14pt
\elevenrm
It is conventionally assumed that the $b$ quark decays left-handedly.  However,
it turns out to be surprisingly hard to exclude the possibility \cite{GW} that
the dominant $b$ decays to charm occur via a {\elevenit right-}handed coupling,
as long as the coupling to leptons in such decays is also right-handed.  One
cannot just look at the beta-decay spectrum, which is the same for $(V-A)
\times (V-A)$ and $(V+A) \times (V+A)$ couplings.  Instead, one needs to study
decays of polarized $b$ quarks.

In the present report, based on the work of Ref.~\cite{ARWW}, we suggest that
the reaction $e^+ e^- \to b \bar b \to \Lambda_b + \ldots$ is likely to give
polarized $b$'s, whose decay to leptons can distinguish among models.  We urge,
in particular, that the LEP experiments (some of which \cite{ALEPH,OPAL} have
already presented evidence for $\Lambda_b$) analyze their data with the
possibility of $(V+A) \times (V+A)$ couplings in mind.  Some aspects of the
polarization studies suggested here have been mentioned previously
\cite{KK,MS}.

The reaction $e^+ e^- \to Z \to b \bar b$ is expected to give rise to $b$
quarks with polarization ${\cal P} \simeq -0.93$ for a weak mixing angle
$\sin^2 \theta = 0.233$.  The $\Lambda_b$ produced in the subsequent
fragmentation of these $b$ quarks should retain this polarization in the
heavy-quark limit [6], in the absence of hard-gluon emission.  The spin of the
$\Lambda_b$ is carried entirely by the $b$ quark; the light quarks in it are
coupled up to spin zero.  If, on the other hand, the $b$ quark ends up in a
(spinless) $B$ meson, information on its polarization is lost.

The inclusive $\Lambda_b \to {\rm charm}$ semileptonic decay can be treated in
the heavy-quark limit as a free-quark decay $b \to c e^- \bar \nu_e$.  If the
$b$ has polarization ${\cal P}$, the expression for the normalized decay
distribution in the $b$ rest frame may be written
$$
\frac{1}{\Gamma} \frac{d^2 \Gamma}{dx~d(\cos \psi)} = \frac{3x^2(1-\zeta)^2}
{f(m_c^2/m_b^2)} \left[ 1 - \frac{2}{3}x + \frac{2x-1}{3}\xi {\cal P} \cos \psi
\right.
$$
$$
\left. +~\zeta \left\{1 - \frac{1}{3}x + \frac{1+x}{3}\xi {\cal P} \cos \psi
\right\} \right]~~,
\eqno(1)
$$
where $x \equiv 2E_e^*/m_b$, $E_e^*$ is the electron energy in the $b$ rest
frame, $\psi$ is the angle between the electron momentum and the spin
quantization axis in this frame, $\xi = \pm 1$ for $V \pm A$ couplings at both
vertices, and
$$
\zeta \equiv \frac{m_c^2}{m_b^2(1-x)}~~~;~~
f(y) \equiv 1 - 8y + 8y^3 - y^4 - 12 y^2 \log y~~~.
\eqno(2)
$$
The electron spectrum (1) is considerably harder for $\xi {\cal P} \cos \psi
\simeq 1$ than for $\xi {\cal P} \cos \psi \simeq -1$, as one can also see
using familiar helicity arguments.

In the reaction $e^+ e^- \to Z \to \Lambda_b + X$, the $b$ quark has a large
momentum along the axis of a fairly well-defined jet. In order to obtain the
signal for $\Lambda_b$ production, one can select events with a lepton at some
minimum transverse momentum $p_T$ with respect to this axis and with an
inclusively produced $\Lambda$ baryon of the appropriate sign [3,4].  In this
frame, let us imagine the $b$ spin to be quantized along its direction of
motion.  The electron transverse momentum and total energy are then $p_T = (m_b
x \sin \psi)/2$ and $E_e = m_b x \gamma_b(1 + v_b \cos \psi) /2$, where $v_b$
is the $b$ quark's velocity and $\gamma_b \equiv (1 - v_b^2)^ {-1/2}$.  Using
these variables we may transform the distribution (1) to obtain
$$
\left. \frac{1}{\Gamma} \frac{d\Gamma}{dE_e} \right|_{p_T \geq p_T^{\rm min}} =
\int dx \theta(p_T - p_T^{\rm min}) \frac {2}{m_b x \gamma_b v_b}
\frac{1}{\Gamma} \frac{d^2 \Gamma}{dx~d(\cos \psi)}~~~
\eqno(3)
$$
Examples of this distribution are shown in Fig.~1 for various values of
$p_T^{\rm min}$ and for the two limiting cases $\xi {\cal P} = 1$ (close to the
standard model) and $\xi {\cal P} = - 1$ (close to the model of Ref.~[1]). We
have taken the $b$ quark to have a laboratory energy of 45 GeV, so that
$\gamma_b = 9$. We have taken $m_c = 1.66$ GeV, $m_b = 5$ GeV, and have ignored
the electron mass.  A distinction on the basis of the electron energy spectrum
is clearly possible, for example, when $p_T^{\rm min} = 0.8$ GeV.

The distributions (1) -- (3) can be convoluted with more realistic functions
for fragmentation of a $b$ quark if desired.  These functions are already in
use in the Monte Carlo simulations employed in Refs.~[3] and [4], so it should
be a simple matter to reproduce distributions analogous to those in Fig.~1
which are more appropriate to the actual experiments.  The fragmentation of a
$b$ quark to a $\Lambda_b$ is not actually known; the corresponding function
for $b \to B$ is peaked around $p_B/p_b \approx 0.7$.  A potential source of
depolarization is production of $\Lambda_b$ via the decay of $\Sigma_b^*$ and
$\Sigma_b$.  If these two resonances are more closely spaced than their natural
widths, they can act coherently to preserve the $b$ quark polarization, but if
they are too widely spaced, some depolarization will result.  (This question
has been considered in Ref.~\cite{bpol}.) The study of the corresponding
spacing between $\Sigma_c$ and $\Sigma_c^*$ should shed some light on this
question.
\vglue 0.5cm
We thank M. Gronau and J. Kroll for helpful comments.  This work was supported
in part by the United States Department of Energy under grant No. DE AC02
90ER40560.
\newpage

\begin{figure}
\vspace{5in}
\caption{Distributions in electron laboratory energy for semileptonic decays of
a $b$ quark with laboratory energy 45 GeV and values of electron $p_T^{\rm min}
= 0,~0.8,$ and 1.6 GeV (curves in descending order). Distributions are
normalized to unit area for $p_T^{min} = 0$.  Solid curves: $\xi {\cal P} =
+1$; dashed curves: $\xi {\cal P} = -1$.}
\end{figure}

\end{document}